\documentclass[prb,twocolumn,tbtags,superscriptaddress,floatfix]{revtex4}
\usepackage{amssymb}
\usepackage{graphicx,amsmath}
\usepackage{bm}
\usepackage{times}

\newcommand{\NLij}{i\kern -0.08em j}

\begin{document}

\title{Dissipative dynamics of two-qubit system: four-level lasing}

\begin{abstract}
The dissipative dynamics of a two-qubit system is studied theoretically. We
make use of the Bloch-Redfield formalism which explicitly includes the
parameter-dependent relaxation rates. We consider the case of two flux
qubits, when the controlling parameters are the partial magnetic fluxes
through the qubits' loops. The strong dependence of the inter-level
relaxation rates on the controlling magnetic fluxes is demonstrated for the
realistic system. This allows us to propose several mechanisms for lasing in
this four-level system.
\end{abstract}

\date{\today }
\author{E.~A.~Temchenko}
\affiliation{B. Verkin Institute for Low Temperature Physics and Engineering, 47 Lenin
Ave., 61103 Kharkov, Ukraine}
\author{S.~N.~Shevchenko}
\email{sshevchenko@ilt.kharkov.ua}
\affiliation{B. Verkin Institute for Low Temperature Physics and Engineering, 47 Lenin
Ave., 61103 Kharkov, Ukraine}
\author{A.~N.~Omelyanchouk}
\affiliation{B. Verkin Institute for Low Temperature Physics and Engineering, 47 Lenin
Ave., 61103 Kharkov, Ukraine}
\maketitle

\section{Introduction}

Recently considerable progress has been made in studying
Josephson-junctions-based superconducting circuits, which can behave as
effectively few-level quantum systems. \cite{Korotkov} When the dynamics of
the system can be described in terms of two levels only, this circuit is
called a qubit. Demonstrations of the energy level quantization and the
quantum coherence provide the basis for both possible practical applications
and for studying fundamental quantum phenomena in systems involving qubits.
Important distinctions of these multi-level artificial quantum systems from
their microscopic counterparts are high level of controllability and
unavoidable coupling to the dissipative environment.

Multi-level systems with solid-state qubits may be realized in different
ways. First, the devices used for qubits in reality are themselves
multi-level systems with the lowest two levels used to form a qubit. For
some recent study of multi-level superconducting devices see Ref. %
\onlinecite{1-qb-multilevel}. Then, a qubit can be coupled to another
quantum system, e.g. a quantum resonator.\cite{qb-oscillator} Such a
composite system is also described by a multi-level structure. As a
particular case of coupling with other systems, the multi-qubit system is of
particular interest (see e.g. Ref. \onlinecite{2qbs}).

Operations with the multi-level systems can be described with
level-population dynamics. In particular, population inversion was proposed
for cooling and lasing with superconducting qubits.\cite{Astafiev07,
Grajcar08-i-drugie} However, most of the previous propositions were related
to three-level systems, while for practical purposes four-level systems are
often more advantageous.\cite{Svelto}

The natural candidate for the solid-state four-level system is the system of
two coupled qubits. The purpose of this paper is the theoretical study of
mechanisms of population inversion and lasing, as a result of the pumping
and relaxation processes in the system. We will start in the next Section by
demonstrating the controllable energy level structure of the system. Our
calculations are done for the parameters of the realistic two-flux-qubit
system studied in Ref.~\onlinecite{Ilichev10}. To describe the dynamics of
the system we will present the Bloch-Redfield formalism in Sec.~III. The key
feature of the system is the strong dependence of the relaxation rates on
the controlling parameters. Then solving the master equation in Sec.~IV we
will demonstrate several mechanisms for creating the population inversion in
our four-level system. We will demonstrate further that applying additional
driving induces transitions between the operating states resulting in
stimulated emission. We summarize our theoretical results in Sec.~V. and,
based on our calculations, we then discuss the experimental feasibility of
the two-qubit lasing.

\section{Model Hamiltonian and Eigenstates of the two-qubit system}

The main object of our study is a system of two coupled qubits. And altough
our analysis bears general character, for concreteness we consider
superconducting flux qubits, see Fig.~\ref{Fig:scheme}. A flux qubit, which
is a superconducting ring with three Josephson junctions, can be controlled
by constant ($\Phi _{\mathrm{dc}}$) and alternating ($\Phi _{\mathrm{ac}%
}\sin \omega t$) external magnetic fluxes. Each of the two qubits can be
considered as a two-level system with the Hamiltonian in the pseudospin
notation \cite{vanderWal03, Korotkov}
\begin{equation}
\widehat{H}_{\mathrm{1q}}^{(i)}=-\frac{1}{2}\epsilon _{i}(t)\widehat{\sigma }%
_{z}^{(i)}-\frac{1}{2}\Delta _{i}\widehat{\sigma }_{x}^{(i)},  \label{H1q}
\end{equation}%
where $\Delta _{i}$ is the tunnelling amplitude, $\widehat{\sigma }%
_{x,z}^{(i)}$ are the Pauli matrices in the basis $\left\{ \lvert {%
\downarrow }\rangle ,\lvert {\uparrow }\rangle \right\} $ of the current
operator in the $i$-th qubit: $\widehat{I}_{i}=-I_{\mathrm{p}}^{(i)}\widehat{%
\sigma }_{z}^{(i)},$ with $I_{\mathrm{p}}^{(i)}$ being the absolute value of
the persistent current in the $i$-th qubit; then the eigenstates of $%
\widehat{\sigma }_{z}$\ correspond to the clockwise ($\widehat{\sigma }%
_{z}\left\vert \downarrow \right\rangle =-\left\vert \downarrow
\right\rangle $) and counterclockwise ($\widehat{\sigma }_{z}\left\vert
\uparrow \right\rangle =\left\vert \uparrow \right\rangle $) current in the $%
i$-th qubit. The energy bias $\epsilon _{i}(t)$ is controlled by constant
and alternating magnetic fluxes
\begin{subequations}
\label{epsilon}
\begin{eqnarray}
\epsilon _{i}(t) &=&2I_{\mathrm{p}}^{(i)}\left( \Phi _{i}(t)-\frac{1}{2}\Phi
_{0}\right) =\epsilon _{i}^{(0)}+\tilde{\epsilon}_{i}(t), \\
\epsilon _{i}^{(0)} &=&2I_{\mathrm{p}}^{(i)}\Phi _{0}f_{i}\text{, \ \ }f_{i}=%
\frac{\Phi _{\mathrm{dc}}^{(i)}}{\Phi _{0}}-\frac{1}{2}, \\
\tilde{\epsilon}_{i}(t) &=&2I_{\mathrm{p}}^{(i)}\Phi _{0}f_{\mathrm{ac}}\sin
\omega t\text{, \ \ }f_{\mathrm{ac}}=\frac{\Phi _{\mathrm{ac}}}{\Phi _{0}}.
\end{eqnarray}

\begin{figure}[h]
\includegraphics[width=8.6cm]{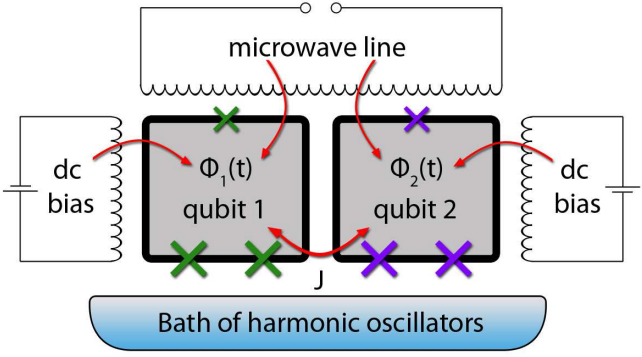}
\caption{(Color online). \textbf{Schematic diagram of the two-qubit system}.
Two different flux qubits are biased by independent constant magnetic
fluxes, $\Phi _{\mathrm{dc}}^{(1)}$ and $\Phi _{\mathrm{dc}}^{(2)}$, and by
the same alternating magnetic flux $\Phi _{\mathrm{ac}}\sin \protect\omega t$%
. The former controls the energy levels structure, while the
latter changes the populations of the levels. The dissipation
processes are described by coupling the system to the bath of
harmonic oscillators.}
\label{Fig:scheme}
\end{figure}


The basis state vectors for the two-qubit system $\left\{ \lvert {\downarrow
\downarrow }\rangle ,\lvert {\downarrow \uparrow }\rangle ,\lvert {\uparrow
\downarrow }\rangle ,\lvert {\uparrow \uparrow }\rangle \right\} $ are
composed from the single-qubit states: $\lvert {\downarrow \uparrow }\rangle
=\lvert {\downarrow }\rangle _{(1)}\lvert {\uparrow }\rangle _{(2)}$, etc.
For identification of the level structure and understanding different
transition rates, we will start the consideration from the case of two
non-interacting qubits. Then, the energy levels of two qubits consist of the
pair-wise summation of single-qubit levels,
\end{subequations}
\begin{equation}
E_{i}^{\pm }=\pm \frac{\Delta E_{i}}{2}=\pm \frac{1}{2}\sqrt{\epsilon
_{i}^{(0)2}+\Delta _{i}^{2}},  \label{DE}
\end{equation}%
which are the eigenstates of the single-qubit time-independent Hamiltonian (%
\ref{H1q}) at $f_{\mathrm{ac}}=0$. We demonstrate this in Fig.~\ref%
{Fig:levels}(a), where we plot the energy levels, fixing the bias in the
first qubit $f_{1}$, as a function of the partial bias in the second qubit $%
f_{2}$. Then the single-qubit energy levels appear as (dashed) horizontal
lines at $E_{1}^{\pm }=\pm \frac{1}{2}\sqrt{\epsilon _{1}^{(0)2}+\Delta
_{1}^{2}}$ for the first qubit and as the parabolas at $E_{2}^{\pm
}(f_{2})=\pm \frac{1}{2}\sqrt{\epsilon _{2}^{(0)}(f_{2})^{2}+\Delta _{2}^{2}}
$.

\begin{figure}[h]
\includegraphics[width=8.6cm]{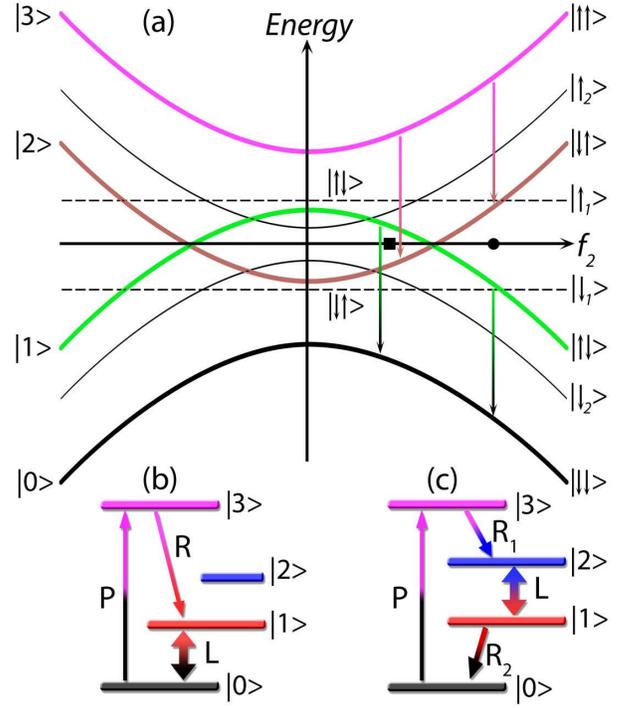}
\caption{(Color online). \textbf{Energy level structure of two uncoupled
qubits} ($J=0$). (a) One-qubit and two-qubits energy levels are shown by
dashed and solid lines as a function of partial flux $f_{\mathrm{2}}$ at
fixed flux $f_{\mathrm{1}}$. We mark the energy levels by the current
operator eigenstates, $\lvert {\downarrow \downarrow }\rangle $ \textit{etc.}
Particularly, we will consider the energy levels and dynamical behaviour of
the system for the flux biases $f_{\mathrm{2}}=f_{2\mathrm{L}}$ (marked by
the square) and $f_{\mathrm{2}}=f_{2\mathrm{R}}$ (marked by the circle). By
the arrows we show the fastest relaxation - for qubit $1$. (b) Scheme for
\textit{three-level lasing} at $f_{\mathrm{2}}=f_{2\mathrm{L}}$. The driving
magnetic flux pumps (P) the upper level $\left\vert 3\right\rangle $. Fast
relaxation (R) creates the population inversion of the first excited level $%
\left\vert 1\right\rangle $ in respect to the ground state $\left\vert
0\right\rangle $; these two operating levels can be used for lasing (L). (c)
Scheme for \textit{four-level lasing} at $f_{\mathrm{2}}=f_{2\mathrm{R}}$.
Pumping (P) and fast relaxations (R$_{1}$ and R$_{2}$) create the population
inversion of the level $\left\vert 2\right\rangle $ with respect to level $%
\left\vert 1\right\rangle $.}
\label{Fig:levels}
\end{figure}

After showing the two-qubit energy levels in Fig.~\ref{Fig:levels}(a), we
assume that the relaxation in the first qubit is much faster than in the
second (this will be studied in the next Section), which is shown with the
arrows in the figure. And now our problem, with four levels and with fast
relaxation between certain levels, becomes similar to the one with lasers.
\cite{Svelto} This allows us to propose three- and four-level lasing schemes
in Fig.~\ref{Fig:levels}(b,c). This is the subject of our further detailed
study.

We have analyzed the relaxation in the system of two uncoupled qubits.
However this system can not be used for lasing, since this requires pumping
from the ground state to the upper excited state (see Fig.~\ref{Fig:levels}%
(b,c)). Such excitation of the two-qubit system requires simultaneously
changing the state of both qubits and can be done provided the two qubits
are interacting. That is why in what follows we consider in detail the
system of two \textit{coupled} qubits. The coupling between the two qubits
we assume to be determined by an Ising-type (inductive interaction) term $%
\frac{J}{2}\widehat{\sigma }_{z}^{(1)}\widehat{\sigma }_{z}^{(2)}$, where $J$
is the coupling energy between the qubits. Then the Hamiltonian of the two
driven flux qubits can be represented as the sum of time-independent and
perturbation Hamiltonians
\begin{gather}
\widehat{H}_{\mathrm{2q}}=\widehat{H}_{0}+\widehat{V}(t),  \label{H2q} \\
\widehat{H}_{0}=\sum_{i=1,2}\left( -\frac{1}{2}\Delta _{i}\widehat{\sigma }%
_{x}^{(i)}-\frac{1}{2}\epsilon _{i}^{(0)}\widehat{\sigma }_{z}^{(i)}\right) +%
\frac{J}{2}\widehat{\sigma }_{z}^{(1)}\widehat{\sigma }_{z}^{(2)},
\label{H0} \\
\widehat{V}(t)=\sum_{i=1,2}-\frac{1}{2}\tilde{\epsilon}_{i}(t)\widehat{%
\sigma }_{z}^{(i)},
\end{gather}%
where $\widehat{\sigma }_{x,z}^{(1)}=\widehat{\sigma }_{x,z}\otimes \widehat{%
\sigma }_{0}$, $\widehat{\sigma }_{x,z}^{(2)}=\widehat{\sigma }_{0}\otimes
\widehat{\sigma }_{x,z}$, and $\widehat{\sigma }_{0}$ is the unit matrix.
When presenting concrete results we will use the parameters of Ref.~%
\onlinecite{Ilichev10}: $\Delta _{\mathrm{1}}/h=15.8$ GHz, $\Delta _{\mathrm{%
2}}/h=3.5$ GHz, $I_{\mathrm{p}}^{(\mathrm{1})}\Phi _{0}/h=375$ GHz, $I_{%
\mathrm{p}}^{(\mathrm{2})}\Phi _{0}/h=700$ GHz, $J/h=3.8$ GHz.

\begin{figure}[h]
\includegraphics[width=8.6cm]{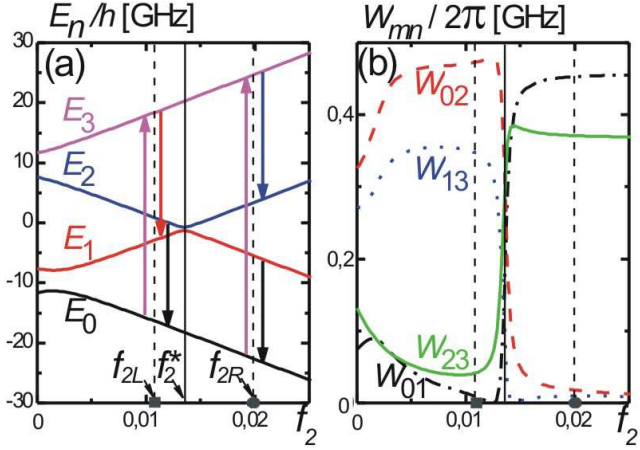}
\caption{(Color online). (a) \textbf{Energy levels} of the system of two
coupled qubits. Arrows show the pumping and dominant relaxation, as in Fig.~%
\protect\ref{Fig:levels}. (b) \textbf{The relaxation rates} $W_{mn}$, which
give the probability of the transition from level $n$ to level $m$, induced
by the interaction with the dissipative bath. Dominant relaxations are $%
W_{13}$ and $W_{02}$ to the left from the avoided crossing at $f_{\mathrm{2}%
}=f_{\mathrm{2}}^{\ast }$ and $W_{23}$ and $W_{01}$ to the right. (The small
relaxation rates $W_{03}$ and $W_{12}$ are not shown.)}
\label{Fig:W1}
\end{figure}

For further analysis of the system, we have to convert to the basis of
eigenstates of the unperturbed Hamiltonian~\eqref{H0}. Eigenstates $\left\{
\lvert {0}\rangle ,\lvert {1}\rangle ,\lvert {2}\rangle ,\lvert {3}\rangle
\right\} $ of the unperturbed Hamiltonian~\eqref{H0} are connected with the
initial basis
\begin{equation}
\left[
\begin{matrix}
\lvert {0}\rangle \\
\lvert {1}\rangle \\
\lvert {2}\rangle \\
\lvert {3}\rangle%
\end{matrix}%
\right] =\widehat{S}\left[
\begin{matrix}
\lvert {\downarrow \downarrow }\rangle \\
\lvert {\downarrow \uparrow }\rangle \\
\lvert {\uparrow \downarrow }\rangle \\
\lvert {\uparrow \uparrow }\rangle%
\end{matrix}%
\right] ,  \label{conversion}
\end{equation}%
where $\widehat{S}$ is the unitary matrix consisting of eigenvectors of the
unperturbed Hamiltonian~\eqref{H0}. Making use of the transformation $%
\widehat{H}_{0}^{\prime }=\widehat{S}^{-1}\widehat{H}_{0}\widehat{S}$, we
obtain the Hamiltonian $\widehat{H}_{0}^{\prime }$ in the energy
representation: $\widehat{H}_{0}^{\prime }=$diag$(E_{0},E_{1},E_{2},E_{3})$.
These eigenvalues of the Hamiltonian $H_{0}$ are computed numerically and
plotted in Fig.~\ref{Fig:W1}(a) as functions of the bias flux in the second
qubit $f_{2}$. The distinction from Fig.~\ref{Fig:levels}(a), calculated
with $J=0$, is in that, first, the crossing at $f_{2}=f_{2}^{\ast }$ becomes
an avoided crossing, and second, the distance between the [previously
single-qubit] energy levels is not equal, e.g. now $E_{3}-E_{2}\neq
E_{1}-E_{0}$.

Likewise, we could also convert the excitation operator $\widehat{V}(t)$ to
the energy representation
\begin{equation}
\widehat{V}^{\prime }(t)=\widehat{S}^{-1}\widehat{V}(t)\widehat{S}%
=\sum_{i=1,2}-\frac{1}{2}\tilde{\epsilon}_{i}(t)\widehat{\tau }_{z}^{(i)},
\label{V'}
\end{equation}%
\begin{equation}
\widehat{\tau }_{z}^{(i)}=\widehat{S}^{-1}\widehat{\sigma }_{z}^{(i)}%
\widehat{S}.
\end{equation}

\section{Master equation and relaxation}

\subsection{Bloch-Redfield formalism}

Following Ref.~\onlinecite{MasterEqn}, we will describe the dissipation in
the open system of two qubits, assuming that it is interacting with the
thermostat (bath), see Fig.~\ref{Fig:scheme}. Within the Bloch-Redfield
formalism, the Liouville equation for the quantum system interacting with
the bath is transformed into the master equation for the reduced system's
density matrix. This transformation is made with several reasonable
assumptions: the interaction with the bath is weak (Born approximation); the
bath is so large that the effect of the system on its state is ignored; the
dynamics of the system depends on its state only at present (Markov
approximation). Then the master equation for the reduced density matrix $%
\rho (t)$ of our driven system in the energy representation can be written
in the form of the following differential equations \cite{MasterEqn}
\begin{equation}
\dot{\rho}_{ij}=-i\omega _{ij}\rho _{ij}-\frac{i}{\hbar }\left[ \widehat{V}%
^{\prime },\widehat{\rho }\right] _{ij}+\delta _{ij}\sum_{n\neq j}\rho
_{nn}W_{jn}-\gamma _{ij}\rho _{ij}.  \label{M_eqn}
\end{equation}%
Here $\omega _{ij}=(E_{i}-E_{j})/\hbar $, and the relaxation rates
\begin{equation}
W_{mn}=2\text{Re}\Gamma _{nmmn},  \label{Ws}
\end{equation}%
\begin{equation}
\gamma _{mn}=\sum_{r}\left( \Gamma _{mrrm}+\Gamma _{nrrn}^{\ast }\right)
-\Gamma _{nnmm}-\Gamma _{mmnn}^{\ast }  \label{gmn}
\end{equation}%
are defined by the relaxation tensor $\Gamma _{lmnk}$, which is given by the
Golden Rule%
\begin{equation}
\Gamma _{lmnk}=\frac{1}{\hbar ^{2}}\int\limits_{0}^{\infty }dte^{-i\omega
_{nk}t}\left\langle H_{\mathrm{I},lm}(t)H_{\mathrm{I},nk}(0)\right\rangle .
\end{equation}%
Here $\widehat{H}_{\mathrm{I}}(t)$ is the Hamiltonian of the interaction of
our system with the bath in the interaction representation; the angular
brackets denote the thermal averaging of the bath degrees of freedom.

It was shown \cite{vanderWal03, Governale01-i-drugie} that the noise from
the electromagnetic circuitry can be described in terms of the impedance $%
Z(\omega )$ from a bath of $LC$ oscillators. For simplicity one assumes that
both qubits are coupled to a common bath of oscillators, then the
Hamiltonian of interaction is written as%
\begin{equation}
\widehat{H}_{\mathrm{I}}=\frac{1}{2}\left( \widehat{\sigma }_{z}^{(1)}+%
\widehat{\sigma }_{z}^{(2)}\right) \widehat{X}  \label{HI}
\end{equation}%
in terms of the collective bath coordinate $\widehat{X}=\sum%
\nolimits_{k}c_{k}\widehat{\Phi }_{k}$. Here $\widehat{\Phi }_{k}$\ stands
for the magnetic flux (generalized coordinate) in the $k$-th oscillator,
which is coupled with the strength $c_{k}$\ to the qubits. We note that the
coupling to the environment in the form of Eq.~(\ref{HI}) applies only to
correlated noise, or both qubits interacting with the same environment. One
could argue that it would be more realistic to use two separate terms, one
for each qubit coupled to its own environment. However, since this term
leads to different relaxation rates in our qubits $1$ and $2$ (see below),
then the form in Eq.~(\ref{HI}) should give essentially the same results as
two separate coupling terms.

Then it follows that the relaxation tensor $\Gamma _{lmnk}$ is defined by
the noise correlation function $S(\omega )$%
\begin{equation}
\Gamma _{lmnk}=\frac{1}{\hbar ^{2}}\Lambda _{lmnk}S(\omega _{nk}),
\end{equation}%
\begin{equation}
\Lambda _{lmnk}=\left( \widehat{\tau }_{z}^{(1)}+\widehat{\tau }%
_{z}^{(2)}\right) _{lm}\left( \widehat{\tau }_{z}^{(1)}+\widehat{\tau }%
_{z}^{(2)}\right) _{nk},
\end{equation}%
\begin{equation}
S(\omega )=\int\limits_{0}^{\infty }dte^{-i\omega t}\left\langle
X(t)X(0)\right\rangle .
\end{equation}%
The noise correlator $S(\omega )$ was calculated in Ref.~%
\onlinecite{Governale01-i-drugie} within the spin-boson model and it was
shown that its imaginary part results only in a small renormalization of the
energy levels and can be neglected. The relevant real part of the relaxation
tensor \cite{Governale01-i-drugie}
\begin{equation}
\text{Re}\Gamma _{lmnk}=\frac{1}{8\hbar }\Lambda _{lmnk}J(\omega _{nk})\left[
\coth \frac{\hbar \omega _{nk}}{2T}-1\right]  \label{ReG}
\end{equation}%
is defined by the environmental spectral density $J(\omega )$. Here $T$ is
the bath temperature ($k_{B}$ is assumed $1$); for the numerical
calculations we take $T/h=1$ GHz ($T=50$ mK). The electromagnetic
environment can be described as an Ohmic resistive shunt across the
junctions of the qubits, $Z(\omega )=R$.\cite{vanderWal03} Then the low
frequency spectral density is linear $J(\omega )\propto \omega Z(\omega
)\propto $ $\omega $ and should be cut off at some large value $\omega _{%
\mathrm{c}}$; the realistic experimental situation is described by \cite%
{Governale01-i-drugie}
\begin{equation}
J(\omega )=\alpha \frac{\hbar \omega }{1+\omega ^{2}/\omega _{\mathrm{c}}^{2}%
},  \label{J(w)}
\end{equation}%
where $\alpha $ is a dimensionless parameter that describes the strength of
the dissipative effects; in numerical calculations we take $\alpha =0.01$
and $\omega _{\mathrm{c}}/2\pi =10^{4}$ GHz (the cut-off frequency $\omega _{%
\mathrm{c}}$ is taken much larger than other characteristic frequencies, so
that for relevant values $\omega :$ $J(\omega )\approx \alpha \hbar \omega $%
).

\subsection{Relaxation rates}

From the above equations the expression for the relaxation rates from level $%
\left\vert n\right\rangle $ to level $\left\vert m\right\rangle $ follows%
\begin{equation}
W_{mn}=\frac{1}{4\hbar }\Lambda _{nmmn}J(\omega _{mn})\left[ \coth \frac{%
\hbar \omega _{mn}}{2T}-1\right] .  \label{Wmn}
\end{equation}%
These relaxation rates are plotted in Fig.~\ref{Fig:W1}(b) as functions of
the partial flux bias $f_{2}$. This figure demonstrates that the fastest
transitions are those between the energy levels corresponding to changing
the state of the first qubit and leaving the same state of the second qubit,
cf. Fig.~\ref{Fig:W1}(a). Namely, the fastest transitions are those with the
rates $W_{13}$ and $W_{02}$ to the left from the avoided crossing and $%
W_{23} $ and $W_{01}$ to the right, which correspond to the transitions $%
\lvert {\uparrow \uparrow }\rangle \rightarrow \lvert {\downarrow \uparrow }%
\rangle $ and $\lvert {\uparrow \downarrow }\rangle \rightarrow \lvert {%
\downarrow \downarrow }\rangle $. Note that we do not show in the figure the
rates $W_{03}$ and $W_{12}$; they correspond to the transitions with
simultaneously changing the states of the two qubits and they are much
smaller than the rates shown.

\begin{figure}[h]
\includegraphics[width=8.6cm]{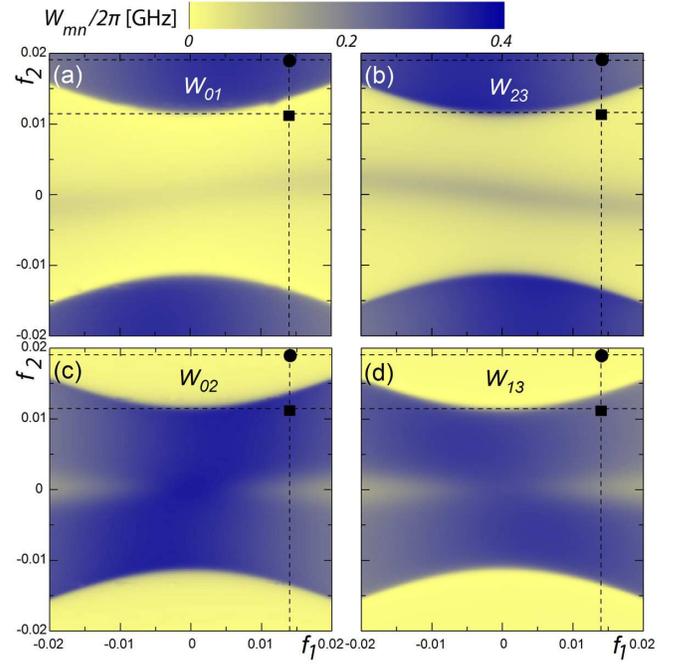}
\caption{(Color online). \textbf{Relaxation rates} $W_{mn}$ versus partial
biases of the two qubits, $f_{\mathrm{1}}$ and $f_{\mathrm{2}}$. The square
and the circle show the parameters $f_{\mathrm{1}}$ and $f_{\mathrm{2}}=f_{%
\mathrm{2L(R)}}$, at which the calculations of other figures are done.}
\label{Fig:W2}
\end{figure}

The relaxation rates $W_{ij}$ are shown in Fig.~\ref{Fig:W2} as functions of
the two partial bias fluxes, $f_{1}$ and $f_{2}$. Again, one can see the
regions where certain relaxation rates are dominant. Such a difference in
the relaxation rates creates a sort of artificial selection rules for the
transitions similar to the selection rules studied in Refs.~%
\onlinecite{Liu05, deGroot}. In our case the transitions are induced by the
interaction with the environment and the difference is due to the different
parameters of the two qubits.\cite{Paladino09} To further understand this
issue, we consider the single-qubit relaxation rates.

From the above equations we can obtain the energy relaxation time $T_{1}$
and the decoherence time $T_{2}$ for single qubit. For the two-level system
with two states $\lvert {0}\rangle $ and $\lvert {1}\rangle $ the relaxation
time is given by \cite{MasterEqn} $T_{1}^{-1}=W_{01}+W_{10}$. The Boltzmann
distribution, $W_{10}/W_{01}=\exp (-\Delta E/T)$, means that at low
temperature the major effect of the bath is the relaxation from the upper
level to the lower one. Now, from Eq.~(\ref{Wmn}) it follows that%
\begin{equation}
T_{1}^{-1}=\frac{\alpha \Delta ^{2}}{2\hbar \Delta E}\coth \frac{\Delta E}{2T%
}.  \label{T1}
\end{equation}%
Also from Eq.~(\ref{gmn}) we obtain the dephasing rate \cite{MasterEqn}%
\begin{equation}
T_{2}^{-1}=\text{Re}\gamma _{01}=\frac{1}{2}T_{1}^{-1}+\frac{\alpha T}{\hbar
}\frac{\epsilon ^{(0)2}}{\Delta E^{2}}.  \label{T2}
\end{equation}%
For the calculation presented in Fig.~\ref{Fig:levels}(a) for two qubits
with $J=0$ in the vicinity of the point $f_{2}=f_{2}^{\ast }$, where $\Delta
E^{(1)}=\Delta E^{(2)}$, we obtain%
\begin{equation}
\frac{T_{1}^{(1)}}{T_{1}^{(2)}}\simeq \left( \frac{\Delta _{2}}{\Delta _{1}}%
\right) ^{2}.  \label{relation}
\end{equation}

As we explained above, the lasing in the four-level system requires the
hierarchy of the relaxation times. In particular, we assumed $T_{1}^{(1)}\ll
T_{1}^{(2)}$. So, in our calculations we have taken $\Delta _{1}\gg \Delta
_{2}$ and consequently the first qubit relaxed faster. This qualitatively
explains the dominant relaxations in Fig.~\ref{Fig:W1}(b).

\subsection{Equations for numerical calculations}

\begin{figure}[h]
\includegraphics[width=8.6cm]{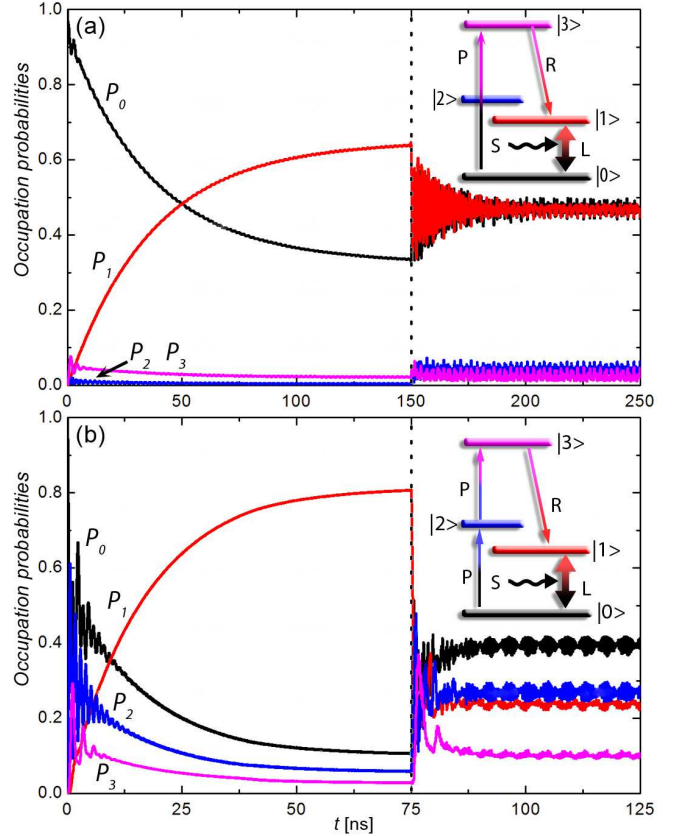}
\caption{(Color online). \textbf{Three-level lasing and stimulated transition%
}. Time evolution of the numerically calculated occupation probabilities at
biases $f_{1}=14\times 10^{-3}$ and $f_{2}=11\times 10^{-3}$ is plotted for
(a) one-photon driving and (b) two-photon driving. As shown in the inset
schemes, the driving and fast relaxation create the inverse population
between the levels $\lvert {1}\rangle $ and $\lvert {0}\rangle $. So, these
levels can be used for lasing, which we schematically mark by the double
arrow. After some time delay (when the population inversion is reached) an
additional periodic signal (S) $f_{\mathrm{L}}\cos \protect\omega _{\mathrm{L%
}}t$ is turned on matching the operating levels, $\hbar \protect\omega _{%
\mathrm{L}}=E_{1}-E_{0}$. This leads to the stimulated transition $\lvert {1}%
\rangle \rightarrow \lvert {0}\rangle $.}
\label{Fig:creation}
\end{figure}

\begin{figure}[h]
\includegraphics[width=8.6cm]{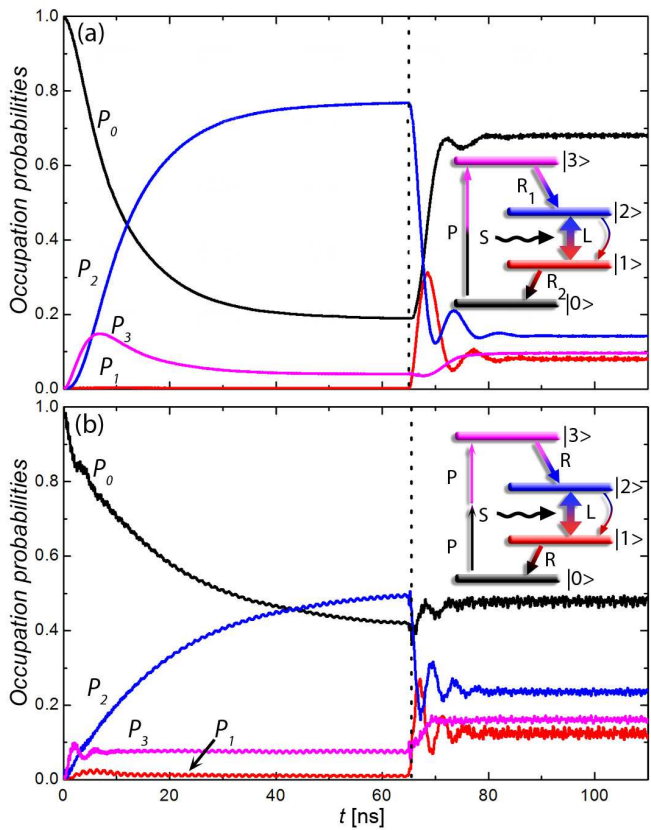}
\caption{(Color online). \textbf{Four-level lasing and stimulated transition}%
. Time evolution of the occupation probabilities at biases $f_{1}=14\times
10^{-3}$ and $f_{2}=20\times 10^{-3}$ is plotted for (a) one-photon driving
and (b) two-photon driving. The driving and fast relaxation create the
inverse population between the levels $\lvert {2}\rangle $ and $\lvert {1}%
\rangle $. After a time delay an additional periodic signal $f_{\mathrm{L}%
}\cos \protect\omega _{\mathrm{L}}t$ is turned on matching the operating
levels, $\hbar \protect\omega _{\mathrm{L}}=E_{2}-E_{1}$. This leads to the
stimulated transition $\lvert {2}\rangle \rightarrow \lvert {1}\rangle $.}
\label{Fig:stimulated}
\end{figure}

If we use the Hermiticity and normalization of the density matrix, then the $%
16$ complex equations~\eqref{M_eqn} can be reduced to $15$ real equations.
After the straightforward parametrization of the density matrix, $\rho
_{ij}=x_{ij}+iy_{ij}$, we get \cite{ShT08}
\begin{subequations}
\label{Eqs}
\begin{gather}
\dot{x}_{ii}=-\frac{1}{\hbar }\left[ V^{\prime },y\right] _{ii}+\sum_{r\neq
i}W_{ir}x_{rr}-x_{ii}\sum_{r\neq i}W_{ii},\text{ }i=1,2,3; \\
\dot{x}_{ij}=\omega _{ij}y_{ij}-\frac{1}{\hbar }\left[ V^{\prime },y\right]
_{ij}-\gamma _{ij}x_{ij},\text{ }i>j; \\
\dot{y}_{ij}=-\omega _{ij}x_{ij}+\frac{1}{\hbar }\left[ V^{\prime },y\right]
_{ij}-\gamma _{ij}y_{ij},\text{ }i>j;
\end{gather}%
$y_{ii}=0$, $x_{00}=1-(x_{11}+x_{22}+x_{33})$; $x_{ji}=x_{ij}$, $%
y_{ji}=-y_{ij}$.

This system of equations can be simplified if the relaxation rates are taken
at zero temperature, $T=0$, and neglecting the impact of the inter-qubit
interaction on relaxation, $J=0$. Then among all the $W_{ij}$ and $\gamma
_{ij}$\ non-trivial are only the elements corresponding to single-qubit
relaxations (see Eqs.~(\ref{T1}-\ref{T2})). For example consider $%
f_{2}<f_{2}^{\ast }$ (see Fig.~\ref{Fig:levels}(a) for the notation of the
levels), then non-trivial elements are
\end{subequations}
\begin{subequations}
\begin{eqnarray}
W_{13} &=&W_{02}=\left( T_{1}^{(1)}\right) ^{-1}=\frac{\alpha \Delta _{1}^{2}%
}{2\hbar \Delta E_{1}}, \\
W_{23} &=&W_{01}=\left( T_{1}^{(2)}\right) ^{-1}=\frac{\alpha \Delta _{2}^{2}%
}{2\hbar \Delta E_{2}},
\end{eqnarray}%
\end{subequations}
\begin{subequations}
\begin{eqnarray}
\gamma _{13} &=&\gamma _{31}=\gamma _{02}=\gamma _{20}=(T_{2}^{(1)})^{-1}=%
\frac{1}{2}(T_{1}^{(1)})^{-1}, \\
\gamma _{23} &=&\gamma _{32}=\gamma _{01}=\gamma _{10}=(T_{2}^{(2)})^{-1}=%
\frac{1}{2}(T_{1}^{(2)})^{-1}.
\end{eqnarray}

In our numerical calculations we did not ignore the influence of the
coupling on relaxation, i.e. we did not assume $J=0$. However, we have
numerically checked that such simplification, $J=0$, resulting in the
relaxation rates (25-26), sometimes allows to describe qualitatively
dynamics of the system.

\section{Several schemes for lasing}

In Sec.~II and in Fig.~\ref{Fig:levels} we pointed out that in the system of
two coupled qubits there are two ways to realize lasing, making use of the
three or four levels to create the population inversion between the
operating levels. In this Section we will demonstrate the lasing in the
two-qubit system solving numerically the Bloch-type equations (\ref{Eqs})
with the relaxation rates given by Eqs.~(\ref{Ws}, \ref{gmn}, \ref{ReG}).
Besides demonstrating the population inversion between the operating levels,
we apply an additional signal with the frequency matching the distance
between the operating levels, to stimulate the transition from the upper
operating level to the lower one. So, we will first consider the system
driven by one monochromatic signal $f(t)=f_{\mathrm{ac}}\sin \omega t$ to
pump the system to the upper level and to demonstrate the population
inversion. Then we will apply another signal stimulating transitions between
the operating laser levels:
\end{subequations}
\begin{equation}
f(t)=f_{\mathrm{ac}}\sin \omega t+f_{\mathrm{L}}\sin \omega _{\mathrm{L}}t.
\end{equation}%
Solving the system of equations (\ref{Eqs}), we obtain the population of $i$%
-th level of our two-qubit system, $P_{i}=x_{ii}$. The results of the
calculations are plotted in Figs.~\ref{Fig:creation} and \ref{Fig:stimulated}%
, where the temporal dynamics of the level populations is presented for
different situations.

In Fig.~\ref{Fig:creation} we consider the situation where the relevant
dynamics includes three levels (for definiteness, we take $f_{1}=14\times
10^{-3}$, $f_{2}=11\times 10^{-3}$, which is marked as the square in Fig.~%
\ref{Fig:W2}). Pumping ($\lvert {0}\rangle \rightarrow \lvert {3}\rangle $)
and relaxation ($\lvert {3}\rangle \rightarrow \lvert {1}\rangle $) create
the population inversion between the levels $\lvert {1}\rangle \ $and $%
\lvert {0}\rangle $. For pumping we consider two possibilities:\ one-photon
driving, Fig.~\ref{Fig:creation}(a), when $\hbar \omega =E_{3}-E_{0}$, and
two-photon driving, Fig.~\ref{Fig:creation}(b), when $2\hbar \omega
=E_{3}-E_{0}$. In the latter case we have chosen the parameters (namely $%
f_{1}$ and $f_{2}$) so, that the two-photon excitation goes via an
intermediate level $\lvert {2}\rangle $. We note here that, as was
demonstrated in Ref.~\onlinecite{Ilichev10}, the multi-photon excitation in
our multi-level system can be direct, as below in Fig.~\ref{Fig:stimulated}%
(b), or ladder-type, via an intermediate level, as in Fig.~\ref{Fig:creation}%
(b). Figure \ref{Fig:creation} was calculated for the following parameters: $%
\omega _{\mathrm{L}}/2\pi =13.7$ GHz ($\hbar \omega _{\mathrm{L}%
}=E_{1}-E_{0} $) and also (a) $\omega /2\pi =35.2$ GHz, $f_{\mathrm{ac}%
}=7\times 10^{-3}$, $f_{\mathrm{L}}=5\times 10^{-3}$; (b) $\omega /2\pi
=17.6 $ GHz, $f_{\mathrm{ac}}=2\times 10^{-3}$, $f_{\mathrm{L}}=5\times
10^{-3}$.

Next, we consider the scheme for the four-level lasing, which occurs in a
similar scenario, except the changing of the levels. Then, the main
relaxation transitions are $\lvert {3}\rangle \rightarrow \lvert {2}\rangle $
and $\lvert {1}\rangle \rightarrow \lvert {0}\rangle $, and now the
population inversion should be created between levels $\lvert {2}\rangle $
and $\lvert {1}\rangle $. For this we take the partial biases $%
f_{1}=14\times 10^{-3}$, $f_{2}=20\times 10^{-3}$ (marked by the circle in
Fig.~\ref{Fig:W2}). First, the system is pumped only with one signal either
with $\hbar \omega =E_{3}-E_{0}$, Fig.~\ref{Fig:stimulated}(a), or with $%
2\hbar \omega =E_{3}-E_{0}$, Fig.~\ref{Fig:stimulated}(b). Such pumping
together with fast relaxation ($\lvert {3}\rangle \rightarrow \lvert {2}%
\rangle $) creates the population inversion between the levels $\lvert {2}%
\rangle \ $and $\lvert {1}\rangle $.\ Fast relaxation from lower laser level
$\lvert {1}\rangle $\ into the ground state $\lvert {0}\rangle $\ helps
creating the population inversion between the laser levels $\lvert {2}%
\rangle \ $and $\lvert {1}\rangle $, which is the advantage of the
four-level scheme.\cite{Svelto} Then the second signal is applied with a
frequency matching the laser operating levels ($\hbar \omega _{\mathrm{L}%
}=E_{2}-E_{1}$). This stimulates the transition $\lvert {2}\rangle
\rightarrow \lvert {1}\rangle $, which provides the scheme for the
four-level lasing. Figure \ref{Fig:stimulated} was calculated for the
following parameters: $\omega _{\mathrm{L}}/2\pi =9$ GHz ($\hbar \omega _{%
\mathrm{L}}=E_{2}-E_{1}$) and also (a) $\omega /2\pi =47.4$ GHz, $f_{\mathrm{%
ac}}=5\times 10^{-3}$, $f_{\mathrm{L}}=3\times 10^{-3}$; (b) $\omega /2\pi
=23.7$ GHz, $f_{\mathrm{ac}}=5\times 10^{-3}$, $f_{\mathrm{L}}=5\times
10^{-3}$.

In the experimental realization of the lasing schemes proposed here, the
system of two qubits should be put in a quantum resonator, e.g. by coupling
to a transmission line resonator, as in Ref.~\onlinecite{Astafiev07}. Then
the stimulated transition between the operating states, which we have
demonstrated here, will result in transmitting the energy from the qubits to
the resonator as photons. For this, the energy difference between the
operating levels should be adjusted to the resonator's frequency.

\section{Conclusions and Discussion}

We have considered the dissipative dynamics of a system of two qubits.
Assuming \textit{different} qubits makes some of the relaxation rates
dominant. With these fast relaxation rates, population inversion can be
created involving three or four levels. The four-level situation is more
advantageous for lasing since the population inversion between the operating
levels can be created more easily. We demonstrated that the upper level can
be pumped by one- or multi-photon excitations. We also have shown that after
applying additional driving, the transition between the operating levels is
stimulated.

When presenting concrete results, we have considered the system of two flux
superconducting qubits with the realistic parameters of Ref. %
\onlinecite{Ilichev10}. For lasing in a generic two-qubit (four-level)
system, our recipe is the following. The hierarchy of the relaxation times
in the system is obtained by making it asymmetric, with different parameters
for individual qubits. This makes transitions between the levels
corresponding to a qubit with smaller tunneling amplitude $\Delta $
negligible, which creates a sort of the artificial selection rule. Based on
our numerical analysis, we conclude that the optimal combination of pumping
and relaxation is realized for $\Delta _{1}\gg \Delta _{2}\sim J$.

Creation of \textit{the population inversion }and\textit{\ the stimulated
transitions} between the laser operating levels, demonstrated here
theoretically, can be the basis for the respective experiments similar to
Ref.~\onlinecite{Astafiev07}. In that work, a three-level qubit (artificial
atom) was coupled to a quantum (transmission line) resonator. First,
spontaneous emission from the upper operating level was demonstrated. In
this way the qubit system can be used as a microwave photon source.\cite%
{Houck09} Then, the operating levels were driven with an additional
frequency and the microwave amplification due to the stimulated emission was
demonstrated. We believe that similar experiments can be done with the
two-qubit system (which forms an \textit{artificial four-level molecule}
from two atoms/qubits). To summarize, we propose to put the two-qubit system
in a quantum resonator with the frequency adjusted with the operating levels
and to measure the spontaneous and stimulated emission as the increase of
the transmission coefficient. Such lasing in a two-qubit system may become a
new useful tool in the qubit toolbox.

\begin{acknowledgments}
We thank E. Il'ichev for fruitful discussions and S. Ashhab for critically
reading the manuscript. This work was partly supported by Fundamental
Researches State Fund (grant F28.2/019) and NAS of Ukraine (project 04/10-N).
\end{acknowledgments}

\end{document}